\documentclass[a4paper]{jpconf}
\usepackage{graphicx}
\usepackage{multirow}
\usepackage{xcolor}
\begin{document}

\title{A new development cycle of the Statistical Toolkit}
\author{M Bati\v{c}$^{1,2}$, A. M. Paganoni$^3$, A. Pfeiffer$^4$, M. G. Pia$^1$, A. Ribon$^4$}

\address{$1$ INFN Sezione di Genova, Genova, Italy}
\address{$2$ Jo\v{z}ef Stefan Institute, Ljubljana, Slovenia}
\address{$3$ MOX - Dept. of Mathematics, Politecnico di Milano, Milano, Italy}
\address{$4$ CERN, Geneva, Switzerland}

\ead{matej.batic@ge.infn.it}

\begin{abstract}
The Statistical Toolkit  is an open source system
specialized in the statistical comparison of distributions. 
It addresses requirements common to different experimental domains, such as
simulation validation (e.g. comparison of experimental and simulated
distributions), regression testing in the course of the software development
process, and detector performance monitoring.
Various sets of statistical tests have been added to the existing
collection to deal with the one sample problem (i.e. the comparison of a
data distribution to a function, including tests for normality,
categorical analysis and the estimate of randomness). 
Improved
algorithms and software design contribute to the robustness of the
results. A simple user layer dealing with primitive data types
%(e.g. parsing comma-separated-values files) 
facilitates the use of the
toolkit both in standalone analyses and in large scale experiments. 
\end{abstract}

\section{Introduction}
The  Statistical Toolkit \cite{Cirrone2004,Mascialino2006} 
was originally conceived as a statistical data analysis
toolkit  for  the problem of comparing data distributions. 
Its development  follows the Unified Software Development Process
\cite{Jacobson1999}. 
According to this approach, the life-cycle of the
software  is iterative-incremental, every iteration representing
an evolution, an improvement, an extension in comparison with the
previous one. 
Iterations in the Statistical Toolkit development process are driven by 
the needs of its experimental applications;
practical use cases steer the implementations of new tests.  

The first development cycles of the Statistical Toolkit implemented a set of
goodness-of-fit (GoF) tests for the two-sample problem, i.e. for the comparison
of two distributions.
These developments were motivated by experimental requirements for regression
testing, validation of simulation with respect to experimental data, comparison
of expected versus reconstructed distributions, and more in general for the
comparison of data from different sources.

New requirements have been identified, based on the experience of
using the Statistical Toolkit in several analyses for the validation of
Geant4 physics models.
%, such as \cite{relax_nist}-\cite{isicsoo}.
These projects highlighted the need for complementary functionality, beyond the 
problem of assessing the compatibility of two distributions.

One of the problems faced in simulation validation (and, more in general, 
in the comparison of experimental distributions), consists in the identification of 
possible systematic effects: tests of randomness address this requirement.

Another problem encountered in the experience with the simulation validation
consists of the comparison not only of individual data distributions, but also
of categories (e.g. the evaluation of differences in the behaviour of two Geant4
physics models with respect to a set of experimental test cases).

\section{Overview of the current functionality of the Statistical Toolkit}

Goodness-of-fit tests quantify the compatibility of the agreement between a set
of sample observations and the the corresponding values predicted from some
model of interest, or between two (or more) sets of observations.
%a random sample with a theoretical distribution function or
%between the empirical distributions of two different populations
%coming from the same theoretical distribution. 
The result of a goodness-of-fit test is expressed through a $p$-value, which
represents the probability that the test statistic has a value at least as
extreme as that observed, assuming the null hypothesis is true.

%The null hypothesis of
%the implemented goodness-of-fit tests is
%\begin{equation}
%H_0: F(x)=G(x),
%\label{null_hypothesis}
%\end{equation}
%while the alternative hypothesis is
%\begin{equation}
%H_0: F(x)\neq G(x).
%\label{alternative_hypothesis}
%\end{equation}
%for the two distributions $F(x)$ and $G(x)$. 

The collection of tests implemented in the current version of the Statistical
Toolkit is given in table \ref{old_tests}; extensive details can be found in
\cite{Cirrone2004,Mascialino2006}.

\begin{table}[htb]
\caption{\label{old_tests}Collection of implemented goodness-of-fit
  tests for comparing two distributions.}
\begin{center}
\begin{tabular}{lll}
\br
GoF test & Distribution Type & \texttt{<ComparisonAlgorithm>} Class \\
\mr
\multirow{4}{*}{Anderson-Darling}
& \multirow{2}{*}{\texttt{Binned}} & \texttt{AndersonDarlingBinned} \\
& & \texttt{AndersonDarlingBinnedApproximated} \\
& \multirow{2}{*}{\texttt{Unbinned}} & \texttt{AndersonDarlingUnbinned} \\
& & \texttt{AndersonDarlingUnbinnedApproximated} \\
\mr
\multirow{3}{*}{Chi-squared}  & \multirow{3}{*}{\texttt{Binned}} 
  & \texttt{Chi2} \\
& & \texttt{Chi2Approximated} \\
& & \texttt{Chi2Integrating} \\
\mr
\multirow{3}{*}{Fisz-Cramer-von-Mises} 
& \texttt{Binned} & \texttt{CramerVonMisesBinned} \\
& \texttt{Unbinned} & \texttt{CramerVonMisesUnbinned} \\
& \texttt{Unbinned} & \texttt{WeightedCramerVonMisesBuningUnbinned} \\
\mr
Girone & \texttt{Unbinned} & \texttt{Girone} \\
\mr
Goodman & \texttt{Unbinned} & \texttt{KolmogorovSmirnovApproximated} \\
\mr
\multirow{3}{*}{Chi-squared}  & \multirow{3}{*}{\texttt{Unbinned}} 
  & \texttt{KolmogorovSmirnov} \\
& & \texttt{WeightedADKolmogorovSmirnov} \\
& & \texttt{WeightedBuningKolmogorovSmirnov} \\
\mr
Kuiper & \texttt{Unbinned} & \texttt{Kuiper} \\
\mr
\multirow{2}{*}{Tiku} 
& \texttt{Binned} & \texttt{TikuBinned} \\
& \texttt{Unbinned} & \texttt{TikuUnbinned} \\
\mr
Watson & \texttt{Unbinned} & \texttt{Watson} \\
\br
\end{tabular}
\end{center}
\end{table}

%\section{New development cycle}

\section{Software improvements}

An effort has been invested to provide an
effective software development environment, which exploits more modern 
tools and facilitates the use of the Statistical Toolkit in a variety of computing
environments.

For the new development cycle Subversion (SVN) \cite{svn} has been selected as a
tool in support of the Configuration and Change Management discipline.
The Statistical Toolkit code was moved to a SVN repository.

In order to facilitate using the Statistical Toolkit on a wide variety of
operating systems, the build system has been moved to the Cross Platform Make
(CMake) \cite{cmake} system.
The \texttt{ctest} testing tool, distributed as a part of CMake, is used for
unit testing.

To be as self-consistent as possible, the number of dependencies on external
software systems has been minimized.
The only essential external dependency is on the GNU Scientific Library
\cite{GSL}.

An additional user layer was implemented to facilitate the use of the
Statistical Toolkit in analysis environment that are concerned neither with AIDA
\cite{Barrand2001} nor with ROOT \cite{Brun1997} analysis objects, which are
supported by the two user layers available in the current version.
The new user layer allows the analyst to supply input data to the Statistical Toolkit
in the form of comma-separated lists of values (csv ASCII files).
If no external dependencies are specified, this user layer is built by default.
Otherwise, in properly set-up environments AIDA or ROOT (or both) 
are found by \texttt{cmake}and the 
corresponding user layer is built automatically.

%The new development cycle of the toolkit has three different user
%layers. 
%Without any external dependencies the user layer that reads
%comma separated list of values  is built by
%default. 
%Additional user layers, to AIDA  and ROOT
%, facilitating usage of Statistical Toolkit using objects from the these
%toolkits (e.g. histograms and ntuples) as data that provides
%distributions. 
%On properly , either or both will be
%when user specifies correct variables.

The Statistical Toolkit comes with an extensive set of \texttt{unitTests},
which are meant to test the correct implementation of the statistical tests for
each new version of the Statistical Toolkit. 

\section{Extension of functionality}

%\begin{figure}[h] \begin{minipage}[t]{0.48\textwidth}
%\includegraphics[width=\textwidth]{diff93_elastic.pdf}
%\caption{\label{relative_differences}Relative difference of the energy deposited
%in the longitudinal slices of the sensitive water volume for various elastic
%scattering configurations, with respect to the reference. Plot taken from
%\protect{\cite{bragg2010}}.} \end{minipage}\hspace{0.02\textwidth}%
%\begin{minipage}[t]{0.48\textwidth}
%\includegraphics[width=\textwidth]{diff9_el.pdf}
%\caption{\label{differences_vs_depth}Differences of the energy deposited for
%various elastic scattering configurations, versus the depth of the sensitive
%water volume.} \end{minipage} \end{figure}

The new development cycle extends the functionality of the Statistical Toolkit
with tests for randomness, one sample goodness-of-fit tests, i.e. comparing data
to reference functions, and tests for categorical data.
Table \ref{new_tests} lists the new tests.

\begin{table}[htb]
\caption{\label{new_tests}Collection of new tests available in the
  latest development version of Statistical Toolkit.}
\begin{center}
\begin{tabular}{lll}
\br
Test & Input data  & Class name \\
\mr
Wald-Wolfowitz 	& Sequence of 	&\texttt{WaldWolfowitzTwoSamplesRunsTest} \\
runs test      	& signs (-1/1)       & \texttt{WaldWolfowitzOneSampleRunsTest} \\
\mr
Wald-Wolfowitz 		& 1-dimensional 	&{\texttt{WaldWolfowitzOneSampleRandomnessTest}} \\
test of randomness 	& distribution 	& \\
\mr
Mann-Whitney $U$ test & 1-dimensional  & \texttt{MannWhitneyTwoSamplesTest} \\
				& distribution & \\
\mr
Fisher's exact test & $2\times2$ matrix & {\texttt{FishersExact2x2Test}} \\
\mr
$\chi^2$ contingency test & $c\times r$ matrix & {\texttt{Chi2ContingencyTableTest}} \\
\mr
$\chi^2$ paired test & Paired values & \texttt{Chi2CurvesComparisonAlgorithm}\\
\br
\end{tabular}
\end{center}
\end{table}

%\section{New tests}
%The new non-parametric tests can be grouped into to categories:
%randomness tests based on runs (sequences of consecutive values of the
%same type e.g. positive or negative numbers) and tests for categorical
%data analysis. 
%
%They typically analyze one sample data, merge the two
%samples into one binary sample (as in case of two sample Wald Wolfowitz
%runs test), or work on categorically grouped observations, as opposed to
%the  goodness of fit tests listed in table \ref{old_tests}. Short descriptions of
%the two groups is presented in following paragraphs.

\subsection{Runs tests of randomness}

Randomness tests provide complementary
information to the existing goodness of fit tests (table \ref{old_tests}): for instance,
tests for randomness can highlight the presence of systematic effects in
the distributions subject to comparison, which goodness of fit  tests cannot
detect. 

A use case is illustrated in \cite{bragg2010}: goodness of fit tests confirm the
compatibility of various Geant4 proton elastic scattering models respect to
reference data, nevertheless asymmetries in the distribution of differences
between the results of the simulation and reference data hint to the presence of
systematic effects associated with some of the Geant4 physics models.

%The development and improvements of the Statistical Toolkit is influenced by
%everyday handling of experimental data and usage of the Statistical Toolkit to
%statistically evaluate it. For instance, during evaluation of Geant4 simulation
%results of proton depth dose profiles to pinpoint the physics-related epistemic
%uncertainties , the distributions of the relative differences of the deposited
%energy for various elastic scattering modeling options compared to the reference
%configuration were comprised within $\pm2$\% (shown in figure
%\ref{relative_differences}) and . Nevertheless, as can be seen from figure
%\ref{differences_vs_depth}, the differences of one of the models with respect to
%the reference configuration clearly

The runs tests are statistical tests, used to test the hypothesis that
the elements of the sequence are mutually independent or whether the
data have some pattern. A run is defined as a series of values of the
same type (e.g. series of increasing/decreasing values, series of
true/false values, etc), the number of consequent values of the same
type being the length of the run.

As an example consider tossing a coin and noting the outcome, which is
either head ($\mathrm{H}$) or tail ($\mathrm{T}$). A run in this example
is each sequence of the same type of outcome. Both too many runs (as in case
of cyclic pattern $\mathrm{HTHT}\ldots$, which has the maximum possible number
of runs for given number of observations) and too few runs (where heads
and tails are clustered together
$\mathrm{HH}\ldots\mathrm{HTT}\ldots\mathrm{T}$) exhibit 
evidence of a non-random relationship between the order of the
experiments and the outcome. 
%The test hence rejects the null hypothesis of
%randomness if either of these outcomes is the case.

The Wald-Wolfowitz test from \cite{Wald1940} is the best known test that
is based on the number of runs. It has been proposed as a test of
whether two samples are from the same population, but as such has 
%very
poor power \cite{Zar2007} and the Mann-Whitney 
%$U$ 
test is preferable.

The new version of the Statistical Toolkit (to be released)
encompasses implementations of the the Wald-Wolfowitz runs test  for one or two
samples. When the test is used with two samples, the algorithm in
\cite{Wald1940} is used to construct one (binary) sample, and results
from the test for one sample are returned. 

To calculate the $p$-values, either the exact or an approximated formula
can be used. The exact calculation of the two-tailed probability of the
test statistics implemented in the Statistical Toolkit follows the description from
\cite{Swed1943}, while the approximated formula takes into account that
for large samples the distribution of the number of runs approaches a
normal distribution. 

Wilcoxon \cite{Wilcoxon1945} published a test for comparison of two
samples, based on comparison of the general size of the two samples,
ranking of the (combined) samples and then comparing the average ranks
of separate ranks. The developments of the test followed fast and the
first to publish it were Mann and Whitney \cite{Mann1947}. The new version of the Statistical Toolkit
implements the Mann-Whitney $U$ test and an approximated formula for the 
$p$-value calculation, again assuming that the samples are large, hence
the distribution of the ranks can be described with a normal
distribution.

\subsection{Tests for categorical data}

%When we separate observations into groups, that share a common trait (a
%nominal attribute, level of scale, numerical value, etc.), we may
%use procedures such as Student's $t$-test to study how continuous random
%variables vary between groups. 

Categorical data analysis involves testing the significance of the association
(contingency) between the groups.
In practice the number of categories 
is usually small (below 20), although in principle the tests for
categorical data could be used for any number of groups. 

The difference between the observed and the expected data, considering
the given marginal and the assumptions of the model of independence, can
be calculated using the $\chi^2$ test (already available in the Statistical Toolkit); 
however, the $\chi^2$ test gives
only an estimate of the true probability value. The estimate might be
inaccurate in case the marginal is very uneven or if there is a small 
value (less than five) in one of the cells of the contingency table.

Fisher's exact test for contingency tables \cite{Fisher1922,Fisher1935}
is most widely known exact test for categorical data analysis. It is
calculated by generating all tables that are more extreme than the table
given by the user. To get the two-tailed $p$-value, the $p$-values of
the tables that have $p$-values of the same size or smaller than the
data table probability are added up to form the cumulative $p$-value,
including the $p$-value of the data table itself. This method becomes
computationally intensive already for moderately sized tables, since the
number of table probabilities to be enumerated can easily reach
billions.  

Fisher's exact test for $2\times2$ contingency tables is available in the 
new development version of the Statistical Toolkit.

An algorithm to calculate the $\chi^2$ test 
with Yates continuity correction has also been implemented as part of the
new development cycle. 

The $\chi^2$ tests can be
applied to general ($c\times r$) contingency tables, while due to
computational reasons Fisher's exact test is only implemented for $2\times 2$ tables. 

%An alternative to Fisher's exact test was published by George Alfred
%Barnard \cite{Barnard1945}, that is now known as Barnard's test. It is
%more time-consuming to compute, but since it can be more powerful
%\cite{Mehta2003} (specially for smaller, that is 2$\times$2 tables), it
%has been added to Statistical Toolkit. {\color{red}Add few sentences about BET.}

\section{Conclusions}
The new development cycle of the Statistical Toolkit comes with a more
versatile build system and provides the user significant
extensions in testing capabilities. 

The new tests extend the Statistical Toolkit 
capabilities with tests for randomness and tests for categorical data
analysis.
The new user layer component makes it possible to use
the Toolkit with many spreadsheet applications that allow exporting
data directly to comma separated list of values.

New tests, together with the new
user layer, make the Statistical Toolkit a powerful data analysis tool
for experimental physics problems concerned with data comparisons.

\section*{Acknowledgment}
This paper is dedicated to the memory of the late Paolo Viarengo, who
contributed to the previous development cycles of the Statistical Toolkit.

\section*{References}

\providecommand{\newblock}{}

%\bibliographystyle{iopart-num}
%\bibliography{bibliography}

\end{document}